\documentclass[twocolumn,aps,showpacs]{revtex4}
\usepackage{bm}
\usepackage{graphicx}
\usepackage{amssymb}
\usepackage{amsmath}
\usepackage{color}
\newcommand{\nix}[1]{}

\usepackage{color}

\begin{document}
\title{Helicity-dependent photocurrents in graphene layers \\
excited by mid-infrared  radiation of a CO$_2$-laser}
\author
{Chongyun Jiang,$^{1,*}$ V.~A.~Shalygin,$^2$ V.~Yu. Panevin,$^2$
S.~N.~Danilov,$^1$ M.~M.~Glazov,$^3$ R.~Yakimova,$^4$
S.~Lara-Avila,$^5$ S.~Kubatkin,$^5$ S.~D.~Ganichev$^{1}$}

\affiliation{$^1$ Terahertz Center, University of Regensburg,
93040 Regensburg, Germany}
\affiliation{$^2$St.\,Petersburg State Polytechnic University, 195251 St.\,Petersburg, Russia}
\affiliation{$^3$ Ioffe Physical-Technical Institute, Russian Academy of Sciences,
194021 St.~Petersburg, Russia}
\affiliation{$^4$
Link{\"o}ping University,
S-58183 Link{\"o}ping, Sweden}
\affiliation{$^5$
Chalmers University of Technology,
S-41296 G{\"o}teborg, Sweden}

\begin{abstract}
We report the study of the helicity driven photocurrents in
graphene  excited by mid-infrared  light of a CO$_2$-laser.
Illuminating an unbiased monolayer sheet of graphene with
circularly polarized radiation generates -- under oblique
incidence -- an electric current perpendicular to the plane of
incidence, whose sign is reversed by switching the radiation
helicity. We show that the current is caused by the interplay of
the circular $ac$ Hall effect and the circular photogalvanic
effect. Studying the frequency dependence of the current in
graphene layers grown on the SiC substrate we observe that the
current exhibits a resonance at frequencies matching the
longitudinal optical phonon in SiC.
\end{abstract}


\date{\today}

\maketitle

\section{Introduction}

Recently graphene has attracted enormous
attention because its unusual electronic properties make possible
relativistic experiments in a solid state environment and may lead
to a large variety of novel electronic
devices~\cite{p1,p2,p2bis,p2bis2}. One of the most interesting
physical aspects of graphene is that its low-energy excitations
are massless, chiral Dirac fermions. The chirality of electrons in
graphene leads to a peculiar modification of the quantum Hall
effect~\cite{p3,p4}, and plays a role in phase-coherent phenomena
such as weak localization~\cite{p5,p6}. Most of current research
in this novel material are focused on the  transport and optical
phenomena. In our recent work, we reported on the observation of
the circular $ac$ Hall effect (CacHE)~\cite{prl2010} which brings
the transport and optical properties of graphene together: In
CacHE an electric current, whose sign is reversed by switching the
radiation helicity, is caused by the crossed electric and magnetic
fields of  terahertz (THz) radiation. The photocurrent is
proportional to the light wavevector and may, therefore, also be
classified as  photon drag
effect~\cite{prl2010,Ch7Barlow54,Ch7Danishevskii70p544,Ch7Gibson70p75,grinberg:class,Ganichevbook,condmat2010}.
Classical theory of CacHE, well describing the experiment at THz
frequencies, predicts that for $\omega \tau \gg 1$, with $\omega$
being the radiation angular frequency and $\tau$ momentum
relaxation time of electrons, the $ac$ Hall effect is suppressed.

Here we demonstrate, however, that helicity driven photocurrents
can be detected applying a  mid-infrared CO$_2$ laser operating at
much higher light frequencies where the condition $\omega \tau \gg
1$ is satisfied. Our results show that in this case, due to the
fact that the classical CacHE is substantially diminished, much
finer effects, such as circular photogalvanic effect (CPGE), well
known for noncentrosymmetric bulk and low dimensional
semiconductors~\cite{Ganichevbook,Ch7Asnin78p74,APL2000,PRB2003,IvchenkoGanichev},
become measurable. We present a phenomenological and microscopic
theory of photocurrents in graphene and show that the experimental
proof of the interplay of CacHE and circular PGE of comparable
strength comes from the spectral behavior of the photocurrent.

Our experiments demonstrate that variation of the radiation
frequency may result in an inversion of the photocurrent sign. We
show that the light frequency, at which the inversion takes place,
changes from sample to sample. Tuning  the radiation frequency in
the operation range of a  mid-infrared CO$_2$ laser we also
observed a resonant-like behaviour of the photocurrent in graphene
grown on the Si-terminated face of a 4H-SiC(0001) substrate: its
amplitude drastically increases at frequency $f = 29.2$~THz
($\lambda =$ 10.26~$\mu$m). The microscopic origin of the resonant
photocurrent is unclear, but we show that its position is
correlated with the high frequency edge of the reststrahlen band
and, correspondingly, to the energy of the LO phonon in 4H-SiC.
Besides the helicity driven electric currents we also present a
detailed study of a photocurrents excited by unpolarized and
linearly polarized light, also observed in our experiments, and
discuss their origin.

\begin{figure}[t]
\includegraphics[width=0.9\linewidth]{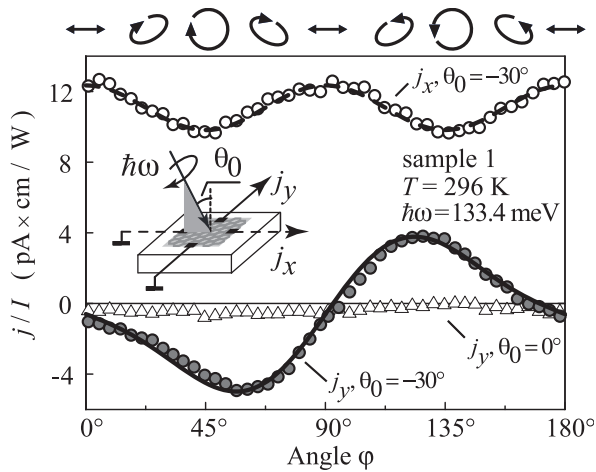}
\caption{Photocurrent $j$ normalized by the light intensity $I$ as
a function of the angle $\varphi$ defining radiation helicity.
Here $j(\varphi)$ is measured at room temperature applying
radiation with $\hbar \omega = 133.4$~meV ($\lambda = 9.27$~$\mu
m$). Open and full circles show the longitudinal, $j_x$, and
transverse, $j_y$, photocurrents measured at oblique incidence
($\theta_0 = -30^\circ$) along and perpendicular to the light
propagation, respectively. Triangles demonstrate that the
photoresponce vanishes at normal incidence ($\theta_0 = 0^\circ$).
Lines show fits according to Eqs.~\eqref{phenom1}, \eqref{jx}
obtained using only photocurrent magnitudes as fitting parameters
[see also Eqs.~\eqref{j:phen}, \eqref{constPD} and \eqref{j:pge},
\eqref{constPGE}]. The inset shows the experimental geometry, the
plane of incidence of the radiation and the arrangement of
contacts (black dots) at the edges of graphene. The ellipses on
top illustrate the polarization states for various $\varphi$ for
light incident on the sample as seen along the propagation
direction. } \label{fig1}
\end{figure}

\begin{figure}[t]
\includegraphics[width=0.8\linewidth]{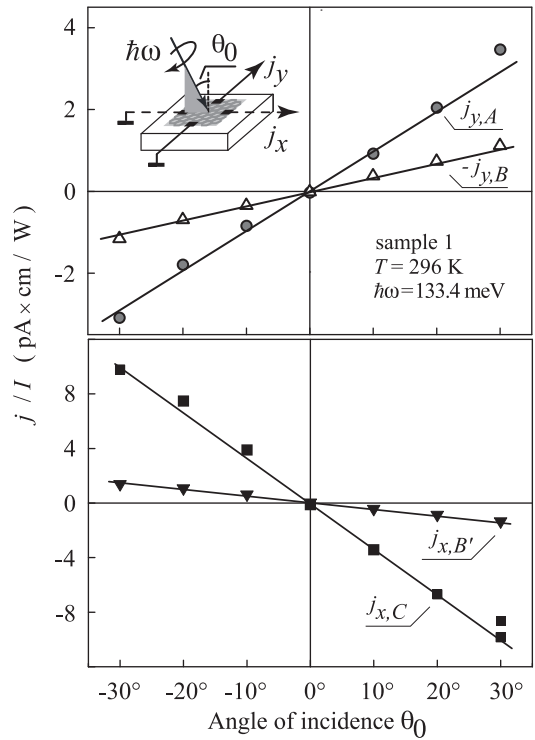}
\caption{ Angle of incidence dependence of the various
photocurrent contributions detected in the transverse (upper
panel) and  longitudinal (lower panel) geometries. Here
$j_{y,A}(\theta_0)$, $j_{y,B}(\theta_0)$, $j_{x,B'}(\theta_0)$,
and $j_{x,C}(\theta_0)$ are obtained by measuring the helicity
dependence of the photocurrent and fitting it by the
Eqs.~\eqref{phenom1} and \eqref{jx} [see also Eqs.~\eqref{j:phen},
\eqref{constPD} and \eqref{j:pge}, \eqref{constPGE}]. The solid
lines are fits after $j \propto \theta_0$. The inset shows the
experiment geometry. } \label{fig2}
\end{figure}

\begin{figure}[t]
\includegraphics[width=0.55\linewidth]{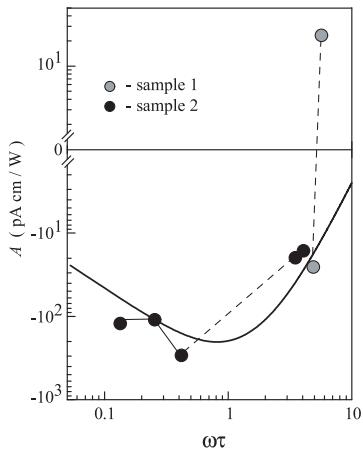}
\caption{ Helicity driven photocurrents given by the coefficient
$A = j_{y,A} /(I \theta_0)$ as function of $\omega\tau$. Solid
curves show calculations of the $ac$ Hall effect. The results of
calculations and the low frequency data ($\omega\tau < 0.4$) are
given after~\cite{prl2010}, see also Eqs.~\eqref{CaCHE}. }
\label{fig2bis}
\end{figure}

\begin{figure}[t]
\includegraphics[width=0.8\linewidth]{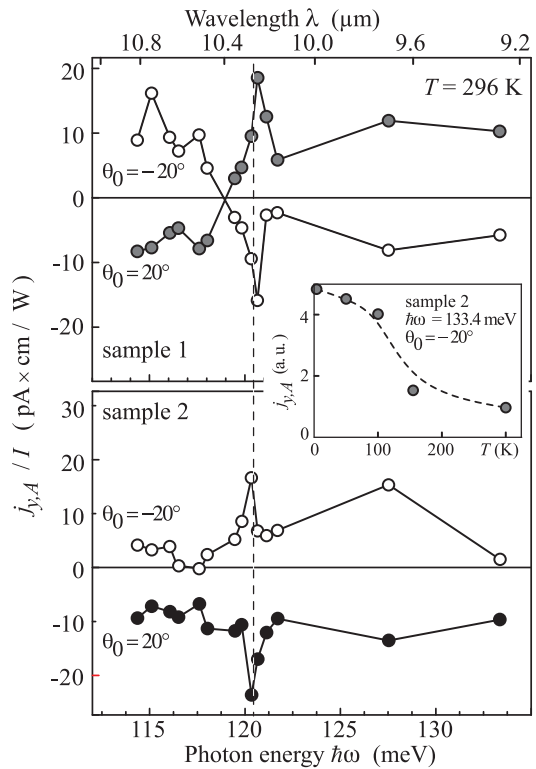}
\caption{Spectral dependence of $j_{y,A}$ obtained for circularly
polarized light ($\varphi = 45^\circ$) and two angles of incidence
$\theta_0 = \pm 30^\circ$. Upper and lower panels show the data
for samples~1 and~2, respectively. The inset shows the temperature
dependence of $j_{y,A}$ measured in sample~2. } \label{fig3}
\end{figure}

\begin{figure}[t]
\includegraphics[width=\linewidth]{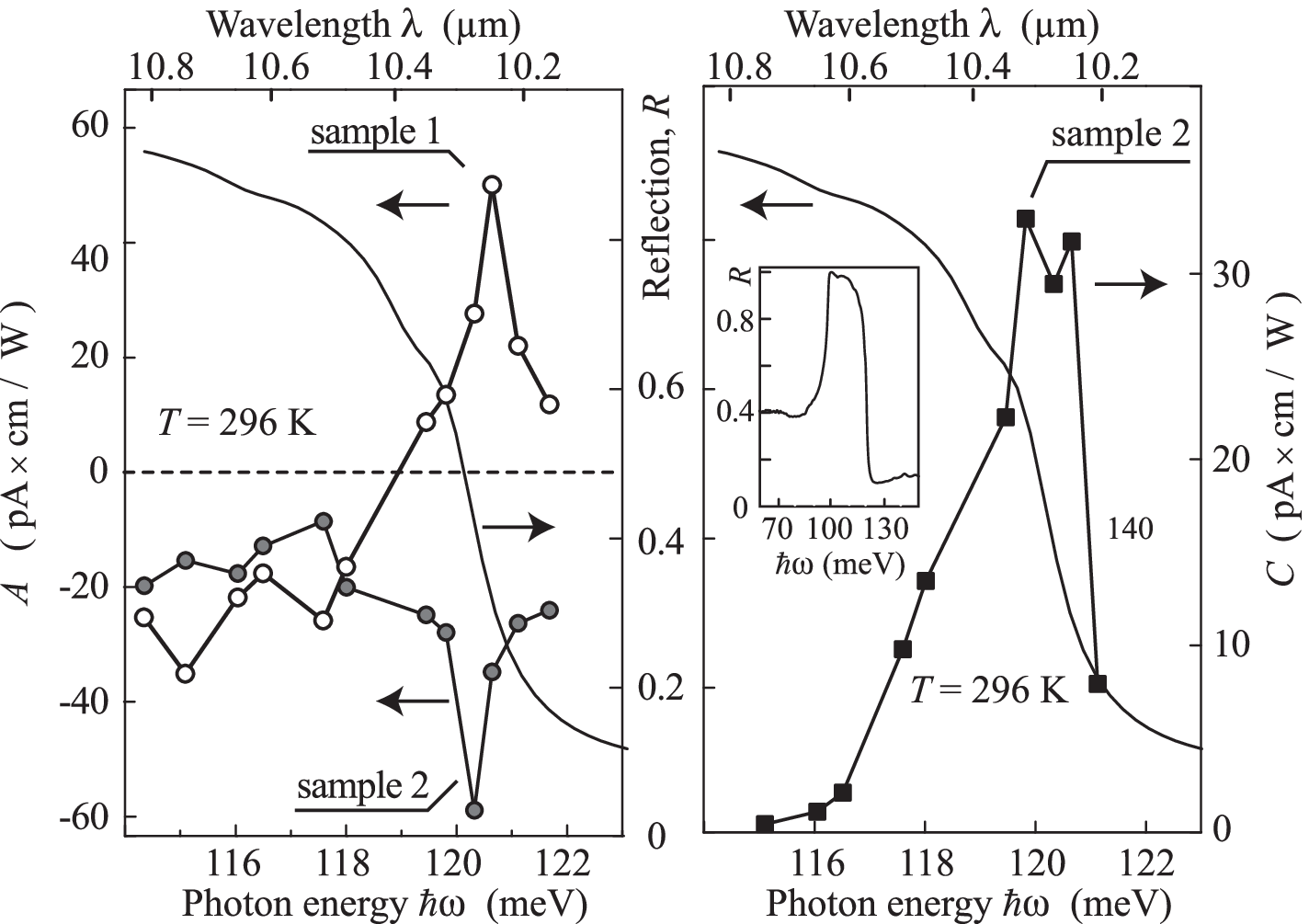}
\caption{Spectral behaviour of  $A=j_{y,A}/(I\theta_0)$ (left
panel) and $C=j_{x,C}/(I\theta_0)$ (right panel) in the vicinity
of resonance. Solid curves show the reflection of the sample. We
note that measurements of the reflection from the graphene and the
back side of the sample yield the almost the same result. The
inset shows the reflection spectrum in a larger frequency range. }
\label{fig4}
\end{figure}

\section{Experiment}

The experiments were  carried out on large area graphene
monolayers prepared by high temperature Si sublimation of
semi-insulating silicon carbide (SiC)
substrates~\cite{LaraAvival09}. The samples have been grown on the
Si-terminated face of a 4H-SiC(0001) substrate. The reaction
kinetics on the Si-face is slower than on the C-face because of
the higher surface energy, which helps homogeneous and well
controlled graphene formation~\cite{Emtsev09}. Graphene was grown
at 2000$^\circ$C and 1\,atm Ar gas pressure resulting in
monolayers of graphene atomically uniform over more than
1000\,$\mu$m$^2$, as shown by low-energy electron
microscopy~\cite{Virojanadara08}. Four contacts have been centered
along opposite  edges of $5 \times 5$~mm$^2$ square shaped samples
by deposition of 3\,nm of Ti and 100\,nm of Au (see inset in
Fig.~\ref{fig1}). The measured resistance was about 2~k$\Omega$.
From low-field Hall measurements, the manufactured material is
$n$-doped due to the charge transfer from
SiC~\cite{Emtsev09,Bostwick09}. We used two layers of
non-conductive polymers~\cite{la11} to protect graphene samples
from the undesired doping in the ambient atmosphere and to control
carrier concentration in the range ($3$ to $7$)$\times$10$^{12}$
cm$^{-2}$,  mobility is of the order of 1000\,cm$^2$/Vs and the
Fermi energies $E_F \sim 300$~meV. All parameters are given for
room temperature.

To generate photocurrents  we applied mid-infrared radiation of
tunable CO$_2$-lasers with operating spectral range from 9.2 to
10.8~$\mu$m (32.6~THz$ \leq f \leq 27.8$~THz) corresponding to
photon energies  ranging from 114 to 135~meV~\cite{Ganichevbook}.
For these wavelengths the conditions $\hbar \omega < E_F$ and
$\omega \tau \gg 1$ hold. Two laser systems were used; a medium
power $Q$-switched laser with the pulse duration of 250~ns
(repetition frequency~of~160~Hz) and low power continuous-wave
($cw$) laser modulated at $120$\,Hz. The samples were illuminated
at oblique incidence with peak power, $P$, of about 500~W and
about 0.1~W for Q-switched and $cw$ laser, respectively. The
radiation power was controlled by photon drag
detector~\cite{Ganichev84p20} and/or MCT detector. The radiation
was focused in a spot of 1~mm diameter being much smaller than the
sample size even at oblique incidence~\cite{exfoliated} . The
initial laser radiation polarization vector was oriented along the
$x$-axis. Applying Fresnel $\lambda/4$ rhomb we modified the laser
light polarization from linear to elliptical. The helicity
$P_{circ}$ of the light at the Fresnel rhomb output was varied
from -1 (left handed circular, $\sigma_-$) to +1  (right handed
circular, $\sigma_+$) according to $P_{circ} = \sin 2\varphi$,
where $\varphi$ is the azimuth of Fresnel's romb. Angle $\varphi =
0$ corresponds to the position of the Fresnel rhomb when its
symmetry plane is oriented perpendicular to the $y$-axis. The
polarization ellipses for some angles $\varphi$ are shown on top
of Fig.~\ref{fig1}.

The geometry of the experiment is sketched in the inset in
Fig.~\ref{fig1}. The incidence angle $\theta_0$ was varied between
$-30^\circ$ and +30$^\circ$. In our experiments we used both
transverse and  longitudinal arrangements in which photoresponse
was probed in directions perpendicular and parallel to the light
incidence plane, respectively (see insets in Fig.~\ref{fig1}
and~\ref{fig2}). The photosignal is measured and recorded with
lock-in technique or with storage oscilloscope. The experiments
were carried in the temperature range from 4.2~K to 300~K.

The signal in unbiased samples is observed under oblique incidence
for both transversal and longitudinal geometries, where the
current is measured in the direction perpendicular and parallel to
the  plane of incidence, respectively. Figure~\ref{fig1} shows the
photocurrent as a function of the angle $\varphi$ for these
geometries. The current behaviour upon variation of radiation
ellipticity is different when measured normal and along to the
light incidence plane.

The photocurrent for the transversal geometry, $j_y$, (see full
circles in Fig.~\ref{fig1}) is dominated by the contribution
proportional to the photon helicity $P_{circ} = \sin 2 \varphi$;
it reverses when the light polarization switches from the
left-handed ($\varphi=45^\circ$) to the right-handed
($\varphi=135^\circ$) light. The overall dependence of $j_y$ on
$\varphi$ is more complex and  well described by
\begin{equation}
\label{phenom1}
j_{y} = j_{y,A} \sin 2\varphi + j_{y,B} \sin 4\varphi + \xi\:,
\end{equation}
where $j_{y,A}=A I \theta_0$ and  $j_{y,B}=B I \theta_0$ are the
magnitudes of the circular and linear contributions, respectively.
Here $I$ is the light intensity. It is noteworthy that the offset
$\xi$ is detected only in some measurements; It is almost zero and
is neglected in the analysis below. The fit to the above equation
is shown in Fig.~\ref{fig1} by solid line. We emphasize, that
exactly the same functional behaviour is obtained from a
phenomenological picture and microscopic models outlined below.
Note that for circularly polarized light, the current is solely
determined by the first term in Eq.~(\ref{phenom1}), because the
degree of linear polarization is zero and, in this case, the
second term vanishes. Our experiments show that $j_{y,A}$ and
$j_{y,B}$ are odd functions of the incidence angle $\theta_0$; a
variation of $\theta_0$ in the plane of incidence changes the sign
of the currents, which vanish for normal incidence, $\theta_0$=0
(see triangles in Fig.~\ref{fig1}). This behaviour is illustrated
by Fig.~\ref{fig2} showing the angle of incidence dependence of
the photocurrents $j_A$, $j_B$ and  $j_C$ determining the
magnitudes of the circular photocurrent and that depending on the
degree of linear polarization, respectively.

In the longitudinal geometry (open circles in Fig.~\ref{fig1}),
the current sign and magnitude are the same for left-handed to
right-handed circular polarized light  and its overall dependence
on $\varphi$ can be well fitted by
\begin{equation}
\label{jx}
j_x = j_{x,B'} \cos 4\varphi + j_{x,C},
\end{equation}
where $j_{x,B'}=B'I \theta_0$ and $j_{x,C}=CI \theta_0$  are the
magnitudes of the linear and polarization-independent contributions,
respectively. The fit after this equation is  shown in Fig.~\ref{fig1} by
dashed line. Like in transversal geometry the photocurrent angular dependence
is in agreement with the theory discussed below.

Figure~\ref{fig2bis} shows spectral behaviour of the circular
photocurrent given by the coefficient $A=j_{y,A}/I\theta_0$. In
this figure $A$ is plotted as a function of $\omega \tau$ for both
graphene samples. Besides the data obtained for light with the
photon energy exceeding 110~meV we included here the results
obtained in the same samples but at much lower THz frequencies $f
\lesssim 4$~THz with $\hbar \omega \lesssim 16$~meV. The latter
data as well as the calculated dependences of the $ac$ Hall effect
are taken from our previous work~\cite{prl2010}. It is seen that
in the second sample the theory of the $ac$ Hall effect describes
well the experiment in the whole frequency range, including the
high frequency data. While the sign and the magnitude of the
current in the first sample measured at low frequency edge of the
CO$_2$-laser operation also fits well to the smooth curve of the
$ac$ Hall effect (see open circles in Fig.~\ref{fig2bis}) at high
frequencies we observed that the signal abruptly changes its sign
with rising frequency. The observed spectral inversion of the
photocurrent's sign reveals that only $ac$ Hall effect can not
describe the experiment.

Figure~\ref{fig3} shows the results of the more detailed study of
the circular photocurrent's frequency dependence. The data were
obtained by using the whole accessible, but very narrow, operating
range of the CO$_2$-laser (114~meV $< \hbar\omega < 135$~meV).
Full and open circles in this figure correspond to the data
obtained for two opposite angles of incidence $\theta_0 = \pm
20^\circ$. It is seen that the detected in sample~1 reversal of
the current direction takes place at $\hbar \omega_{inv} \simeq
119$~meV. Here, $\omega_{inv}$ indicates the frequency of the sign
inversion. The drastic difference in  the photocurrent's spectral
behaviour detected for samples with almost the same mobility and
carrier density but prepared not in the same growth circle we
attribute to the change of coupling between graphene layer and the
substrate. In fact, this parameter is crucial for the mechanisms
of the photocurrent generation. It may be different from sample to
sample and it is difficult to control.

Besides the spectral inversion, we observe another remarkable
feature of the photocurrent: in both samples we detected a
resonance increase of the current magnitude at $\hbar \omega
\simeq 121$~meV (see Fig.~\ref{fig3} and left panel
in~Fig.~\ref{fig4}). Similar resonance-like behaviour is detected
for the polarization-independent longitudinal photocurrent (see
right panel in Fig.~\ref{fig4}). The position of the resonance
corresponds to the longitudinal optical (LO) phonon energy in
4H-SiC. In order to prove this we measured the sample reflection
for the graphene and the substrate sides. The results for both
sides  almost coincide with each other: the reflection shows the
reststrahlen band behaviour (see the inset in Fig.~\ref{fig4}).
Solid curves in the left and right panels in Fig.~\ref{fig4} show
that the high frequency edge of the reststrahlen band, which
corresponds to the LO phonon energy in 4H-SiC, coincide with the
resonance position.  The
detailed study of the resonance photocurrent and its power
dependence is beyond the scope of the present work.

To summarize the experimental part we demonstrate that
illumination of graphene monolayers by mid-infrared radiation at
oblique incidence results in the generation of photocurrents.
Their directions and magnitudes are determined by the polarization
of the radiation. At the frequencies about $f=29.2$~THz ($\hbar
\omega = 121$~meV) we observed resonance feature and sign
inversion of the photocurrent. The latter property is sample
dependent.

\section{Theory}

Below we present phenomenological analysis of the
photocurrents in graphene as well as their microscopic models. We
demonstrate that the experimentally observed incidence angle,
linear polarization and helicity dependences of the photocurrents correspond to
phenomenological models. The magnitudes of the photocurrents and their polarization
dependencies are also in good agreement with theoretical predictions.

\subsection{Phenomenological analysis}

The ideal honeycomb lattice of graphene is described by the point
group $D_{6\rm h}$ containing the spatial inversion. As a result,
photocurrent generation is possible provided that the joint action
of electric, $\bm E$, and magnetic, $\bm B$, fields of the
radiation is taken into account or provided that the allowance for
the radiation wave vector, $\bm q$, transfer to electron ensemble
is made. In the former case the Cartesian components of the
current are proportional to the bi-linear combinations $E_\alpha
B_\beta^*$, while in the latter case to the combinations $q_\alpha
E_\beta E_\gamma^*$. Here Greek subscripts enumerate Cartesian
components. For the plane wave its wave vector, electric and
magnetic fields are interrelated, therefore, for the purposes of
the phenomenological analysis it is enough to express the
photocurrent density via the combinations $q_\alpha E_\beta
E_\gamma^*$  as~\cite{condmat2010}
\begin{subequations}
\label{j:phen}
\begin{equation}
\label{j:x}
j_x/I = T_1  q_x \frac{|e_x|^2 + |e_y|^2}{2}
+ T_2 q_x \frac{|e_x|^2 - |e_y|^2}{2}  ,
\end{equation}
\begin{equation}
\label{j:y}
j_y / I = T_2 q_x \frac{e_xe_y^* + e_x^*e_y}{2}  - \tilde{T}_1  q_x P_{\rm circ} \hat{e}_z .
\end{equation}
\end{subequations}
where $x$ and $y$ are the axes in the graphene plane, and $z$ is
the structure normal, the radiation is assumed to be incident in
$(xz)$ plane, $\hat{\bm e}$ is the unit vector in light
propagation direction and $\bm e$ is the (complex) polarization
vector of radiation, $P_{\rm circ}$ is the circular polarization
degree and $\bm q$ is the radiation wave vector. Additional
contributions to the photocurrents, involving $z$ component of
electric field are analyzed in Ref.~\cite{condmat2010}. These
effects are expected to be strongly suppressed in \textit{ideal}
samples and for moderate radiation frequencies.
Expressions~\eqref{j:phen} can be rewritten via incidence angle,
$\theta_0$, and angle $\varphi$ determining the radiation helicity
as Eqs.~\eqref{phenom1}, \eqref{jx}. It allows one to establish a
link between phenomenological constants $T_1$, $T_2$ and $\tilde
T_1$ and fitting parameters $A$, $B$ and $C$ used to describe the
experimental data, see Figs.~\ref{fig1} and \ref{fig2}. Namely, at
small incidence angles
\begin{equation}
\label{constPD}
 A \propto \tilde T_1, \quad B \propto T_2, \quad \mbox{and} \quad C \propto T_1.
\end{equation}

\begin{figure}[t]
\includegraphics[width=\linewidth]{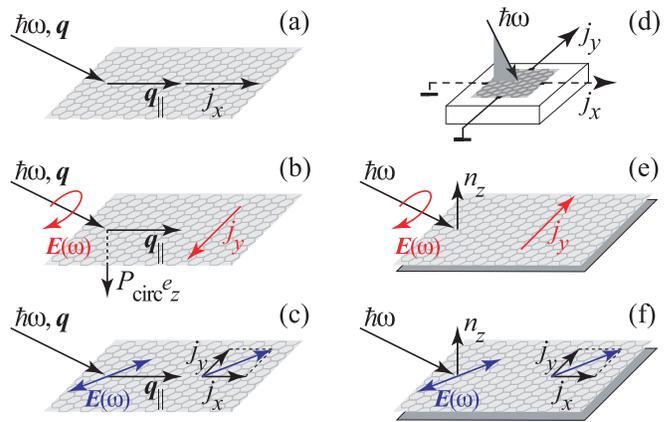}
\caption{Schematic illustration of the possible contributions
to the photon drag and photogalvanic effects.
Panels (a)-(c): polarization independent, circular and linear
photon drag effects, Eqs.~\eqref{j:phen}. Panel (d) shows the experimental geometry.
Panels (e)-(f):  photogalvanic effects allowed by symmetry
in graphene samples deposited on substrates.
} \label{figt1}
\end{figure}

It follows from Eqs.~\eqref{j:phen} that photocurrent contains, in
general, three contributions illustrated in Fig.~\ref{figt1}, panels (a)--(c).
First one, schematically illustrated in Fig.~\ref{figt1}(a) results in the
polarization-independent photocurrent flowing along the light incidence plane.

In accordance with the general line of the paper we pay special
attention to the photocurrent contribution  presented in
Fig.~\ref{figt1}(b) where the generation of the transversal to the
light incidence plane current is shown. This current component is
dependent on the radiation helicity: by changing photon from
right- to left- circularly polarized, current changes its
direction. This is nothing but the CacHE uncovered recently in
graphene Ref.~\cite{prl2010}. In addition, transversal
photoresponse contains a component, being sensitive to the linear
polarization of radiation, see Fig.~\ref{figt1}(c).

The photocurrent components described by Eqs.~\eqref{j:phen} can also be
qualified as  photon drag effects~\cite{Ganichevbook,ivchenkobook} since in their
phenomenological description the photon wave vector is involved.
The direction of the photocurrent changes its sign upon reversal
of the incidence angle.
The contributions given by Eq.~\eqref{j:x} and the first term
in the Eq.~\eqref{j:y} can be easily understood as transfer of linear
momenta of photons to the electron system~\cite{JETP1982} and is
recently discussed for graphene~\cite{condmat2010,Entin:gr}.
The circular photon drag current described by the
second term on the right hand side of Eq.~\eqref{j:y} is due to transfer of both
linear and angular momenta of photons to free carriers. The circular
photon drag effect was discussed
phenomenologically~\cite{Ivchenko1980,Belinicher1981} and observed in
GaAs quantum wells in the mid-infrared range~\cite{Shalygin2006} and
in metallic photonic crystal slabs~\cite{Hatano09}.
We note, that while the microscopic description of the circular photocurrent in graphene
in terms of $ac$ Hall effect is relevant
to the relatively low radiation frequencies range at high frequencies
all photocurrent contributions can be conveniently treated
in terms of photon drag effect.

The real structures, however, are deposited on a substrate, which
removes the equivalence of the $z$ and $-z$ directions and reduces
the symmetry to the $C_{6\rm v}$ point group. Such symmetry reduction makes
photogalvanic effects possible. The photogalvanic effects give rise to the
linear and circular photocurrents~\cite{condmat2010}:
\begin{subequations}
\label{j:pge}
\begin{equation}
\label{j:pge:x}
j_{x}/I = \chi_{l} \frac{e_x e_z^*+ e_x^*e_z}{2} \:,
\end{equation}
\begin{equation}
\label{j:pge:y}
j_{y}/I = \chi_{l} \frac{e_y e_z^*+ e_y^*e_z}{2} +
\chi_{c} P_{\rm circ} \hat{ e}_x  \:,
\end{equation}
\end{subequations}
described by two independent parameters $\chi_l$ and $\chi_c$.
Schematically, these contributions to the photocurrent are shown in
Fig.~\ref{figt1}(e), (f). It follows from Eqs.~\eqref{j:pge} that the
linear photocurrent flows along the projection of the electric field
onto the sample plane and it has both $x$ and $y$ components, in
general. By contrast, circular photocurrent flows transverse to the
radiation incidence plane, i.e. along $y$ axis in the chosen
geometry. Despite the fact that the photogalvanic effects
described by Eqs.~\eqref{j:pge} require the out-of-plane component
of the incident radiation, they may be important for real graphene
samples as it follows from the microscopic model, see Sec.~\ref{micro}.

Equations~\eqref{j:pge} reveal that transverse and longitudinal
photogalvanic currents vary upon change of the radiation polarization similarly to
the $ac$ Hall (photon-drag) photocurrents given by Eqs.~\eqref{j:phen}.

Thus,  photogalvanic effects described by Eqs.~\eqref{j:pge} make
only additional contributions to the constants $A$, $B$ and $C$ in
phenomenological expressions~\eqref{phenom1}, \eqref{jx}. In the
case that the photocurrent is driven solely by photogalvanic
effects these constants are given by,
\begin{equation}
\label{constPGE}
 A \propto \chi_c, \quad \mbox{and} \quad  -2B = C  \propto \chi_l.
\end{equation}
It follows from Eqs.~\eqref{j:phen} and \eqref{j:pge} that
the phenomenological theory, which is based solely on symmetry
arguments and does not require knowledge of the microscopic
processes of light-matter coupling in graphene, describes well the
polarization dependences of the photocurrents presented in
Fig.~\ref{fig1} and fitted by Eqs.~\eqref{phenom1} and
\eqref{jx}. The incidence angle dependences presented in Fig.~\eqref{fig2}
are also in line with phenomenological description.

Hence, the phenomenological analysis is presented, which yields a
good agreement with the experiment. Schematical illustration
Fig.~\ref{figt1} as well as Eqs.~\eqref{j:phen} and \eqref{j:pge}
show that both the $ac$ Hall effect and photogalvanic effect have
almost the same polarization and incidence angle dependences.
Therefore, the analysis of polarization and incidence angle
dependencies of the photocurrents is not enough to establish their
microscopic origins. Therefore, extra arguments based on
microscopic model are needed.

\subsection{Microscopic mechanisms}\label{micro}

Before turning to the presentation  of the microscopic models, let
us introduce the different regimes of radiation interaction with
electron ensemble in graphene depending on the photon frequency,
$\omega$, electron characteristic energy (Fermi energy), $E_F$,
and its momentum relaxation rate $1/\tau$. We assume that the
condition $E_F\tau/\hbar \gg 1$ is fulfilled (which is the case
for the samples under study) making possible to treat electrons in
graphene as free.

If photon energy is much smaller  compared with electron Fermi
energy, $\hbar\omega \ll E_F$, the classical regime is realized.
In this case the electron motion can be described within the
kinetic equation for the time $t$, momentum $\bm p$ and position
$\bm r$ dependent distribution function $f(\bm p, \bm r, t)$.

An increase of the photon energy makes classical approach invalid.
If $\hbar\omega \leqslant 2E_F$ the direct interband transitions
are not possible and the radiation absorption as well as the
photocurrent generation are possible via indirect (Drude-like)
transitions. It is worth to mention that if $\hbar/\tau \ll
\hbar\omega \leqslant E_F$ the transitions are intraband, while
for $E_F<\hbar\omega \leqslant 2 E_F$ the initial state for the
optical transition may be in the valence band.

In what follows we restrict  ourselves to the indirect intraband
transitions, assuming that $\hbar/\tau \ll \hbar\omega \leqslant
E_F$ which corresponds to our experiments with CO$_2$ laser
excitation. The results for the classical frequency range,
$\hbar\omega\ll E_F$, relevant for THz excitation, will be also
briefly discussed.

\subsubsection{High frequency (\emph{ac}) Hall effect}

The microscopic calculation of the \emph{ac} Hall effect
in the classical frequency range, where $\hbar\omega \ll E_F$ was
carried out in Refs.~\cite{prl2010,condmat2010}.
Thus, we give here only the final result of this work obtained
within the framework of the Boltzmann equation with allowance for
both $\bm E \bm B$ (\emph{ac} Hall effect) and $qE^2$ (spatial
dispersion effect) contributions. The circular photocurrent is
given for degenerate electrons by
%
\begin{multline}
\label{CaCHE} j_A = A \theta_0 \sin{2\varphi} = \\
 = q \theta_0
\frac{e^3 {\tau_1 (v \tau_1 E)^2 }}{{2 \pi \hbar^2 (1 + \omega^2
\tau_1^2)}} P_{\rm circ} \left( 1 + \frac{\tau_2}{\tau_1}\right)
\frac{1 - r}{1 + \omega^2 \tau_2^2},
\end{multline}
Here $q = \omega / c$, we have replaced for the small
incidence angles $q \sin{\theta_0} \approx q \theta_0$, $v$
  is the electron velocity in graphene, $\tau_1$ and
$\tau_2$ are the relaxation times of first and second angular
harmonics of the distribution function describing the decay of the
electron momentum and  momentum
alignment~\cite{prl2010,condmat2010,perel73}, and $r = d{\rm
ln}\tau_1/d{\rm ln}\varepsilon$ ($\varepsilon$ is the electron
energy). The frequency dependence is presented by a solid curve in
Fig.~\ref{fig2bis}. At low frequencies $\omega\tau \ll 1$ the
parameter $A$ and, correspondingly, the circular photocurrent
raises with the  frequency increase as $\omega\tau$.
In the high  frequency regime $\omega\tau\gg 1$, by contrast, the
circular photocurrent related with CacHE drops as
\begin{equation}
\label{hfA}
 j_A \propto \frac{1}{\omega^3\tau}, \quad \frac{\hbar}{\tau} \ll \hbar \omega \ll E_F.
\end{equation}
Calculations show that for our $n$-type structures the constant
$A$ describing CacHE photocurrent is negative in the wholef
frequency range, achieves its maximum absolute value for
$\omega\tau \sim 1$ and describes well the experiment at least at
low frequencies (see Figure~\ref{fig2bis}).

The solution of the Boltzmann equation also yields linear
photocurrents in longitudinal ($j_{B^\prime}$ and $j_C$) and
transverse ($j_{B}$) geometries~\cite{prl2010,condmat2010}. These
photocurrents are proportional to the constants $T_1$ and $T_2$ in
Eqs.~\eqref{j:phen}. They describe well polarization dependences
presented in Fig.~\ref{fig1} providing the polarization
independent longitudinal photocurrent as well as photocurrent
contributions varying with the change of degree of linear
polarization as $\sin 4\varphi$ and $\cos 4\varphi$. These
constants $T_1$ and $T_2$ as functions of frequency diverge as
$1/\omega$ at $\omega\tau\to 0$ and decay as $1/\omega^3$  for
$\omega\tau\gg 1$.  As a result,
\begin{equation}
\label{hfBC}
 j_B, j_C \propto \frac{1}{\omega^2}, \quad \frac{\hbar}{\tau} \ll \hbar \omega \ll E_F.
\end{equation}
It should be noted that the longitudinal linear photocurrent can
change its direction as function of the radiation frequency depending
on the dominant scattering mechanism~\cite{condmat2010}.

\begin{figure}[hptb]
\includegraphics[width=\linewidth]{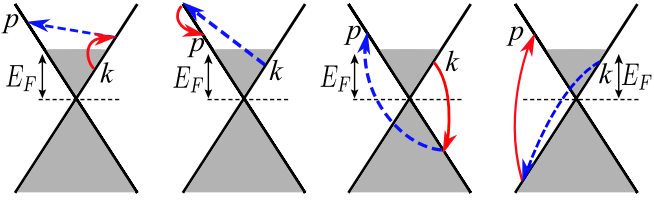}
\caption{Schematic illustration of the processes responsible for the
  drag effect in the quantum frequency range under intraband
  transitions ($\hbar\omega \leqslant E_F$). Solid/red arrows denote electron-photon interaction,
  dashed/blue arrows denote electron scattering caused by impurities or phonons.
  Filled/gray area shows the part of energy spectrum filled with electrons. }
\label{fig:inter}
\end{figure}

To present a complete picture of the photocurrent formation due to
Drude absorption we turn to the quantum frequency range and assume
that $\hbar\omega \leqslant E_F$, while $\omega\tau \gg 1$. The
absorption of the electromagnetic wave in the case of intraband
transitions should be accompanied with the electron scattering,
otherwise energy and momentum conservation laws can not be
satisfied. The matrix elements describing electron transition from
$\bm k$ to $\bm p$ state with the absorption ($ M_{\bm p, \bm
k}^{{\rm abs},\bm q}$) and emission ($M_{\bm p, \bm k}^{{\rm
emit},\bm q}$) of a photon with the wave vector $\bm q$ are
calculated in the second order of perturbation theory as
\begin{subequations}
\label{m:second}
\begin{equation}
 M_{\bm p, \bm k}^{{\rm abs},\bm q} = \sum_{\nu=\pm} \left\{ \frac{ V^{+\nu}_{\bm p, \bm k
 +\bm q} R^{\nu+}_{\bm k+\bm q,\bm k} }{ \varepsilon_{\bm k}^{+} -\varepsilon_{\bm k+\bm q}^\nu
 + \hbar \omega } + \frac{ R^{+\nu}_{\bm p, \bm p -\bm q} V^{\nu+}_{\bm p-\bm q,\bm k} }{ \varepsilon_{\bm k}^{+}
 -\varepsilon_{\bm p-\bm q}^\nu  }\right\}, \label{M:abs}\\
\end{equation}
\begin{equation}
M_{\bm p, \bm k}^{{\rm emit},\bm q} = \sum_{\nu=\pm} \left\{ \frac{ V^{+\nu}_{\bm p, \bm k
-\bm q} R^{\nu+}_{\bm k-\bm q,\bm k} }{ \varepsilon_{\bm k}^{+} -\varepsilon_{\bm k-\bm q}^\nu
- \hbar \omega } + \frac{ R^{+\nu}_{\bm p, \bm p+\bm q} V^{\nu+}_{\bm p+\bm q,\bm k} }{ \varepsilon_{\bm k}^{+}
-\varepsilon_{\bm p+\bm q}^\nu  }\right\}. \label{M:emit}
\end{equation}
\end{subequations}
Here superscript $\nu$ enumerates conduction band ($\nu=+$) and
valence band ($\nu=-$), respectively, $R_{\bm k\pm \bm q, \bm
k}^{\nu\nu'}$ is the electron-photon interaction matrix element,
$V_{\bm p, \bm k}^{\nu\nu'}$ is the matrix element describing
electron scattering by an impurity or a phonon. We note that the
incident electromagnetic wave is assumed to be classical, hence
the electron-photon interaction matrix elements are the same for
the emission and absorption processes, $M_{\bm p, \bm k}^{{\rm
emit},\bm q} = M_{\bm k, \bm p}^{{\rm abs},\bm q} \equiv M_{\bm p,
\bm k}^{\bm q}$ because the number of photons in this wave is
large. It was assumed also that the graphene is $n$ doped so the
initial and final states lie in the conduction band. The
intermediate state, however, can be in conduction or in valence
bands, see Fig.~\ref{fig:inter}.

The \textit{dc} current density can be calculated
as~\cite{grinberg:quant}
\begin{multline}
 \label{j:quant}
\bm j =
e\frac{8\pi}{\hbar}\sum_{\bm k, \bm p} [\bm v_{\bm p} \tau_1(\varepsilon_{ p})
- \bm v_{\bm k} \tau_1(\varepsilon_{ k})] |M_{\bm p, \bm k}^{\bm q}|^2 \\
[f(\varepsilon_k) - f(\varepsilon_p)] \delta(\varepsilon_p- \varepsilon_k -\hbar\omega),
\end{multline}
where $\bm v_{\bm k}$ is the electron velocity in the state with
the wave vector $\bm k$, $\tau_1(\varepsilon_k)$ is the momentum
relaxation time, $f(\varepsilon_k)$ is the Fermi-Dirac
distribution function, $\varepsilon_k=\hbar v k$ is the electron
dispersion in graphene.

Let us assume that the electron scattering is provided by the
short-range impurities acting within given valley, intervalley
scattering processes are disregarded. The matrix elements for the
impurity scattering are given by
\[
V^{++}_{\bm p \bm k} = \frac{V_0}{2} \left[1+ e^{\mathrm
    i(\varphi_{\bm k} - \varphi_{\bm p})}\right], \quad V^{-+}_{\bm p \bm k} = \frac{V_0}{2} \left[1- e^{\mathrm
    i(\varphi_{\bm k} - \varphi_{\bm p})}\right],
\]
\begin{equation}
\label{V++}
 V^{+-}_{\bm p \bm k} = \frac{V_0}{2} \left[1- e^{\mathrm
    i(\varphi_{\bm k} - \varphi_{\bm p})}\right],
\end{equation}
where $V_0$ is real constant.
 As a result, one can express the coefficients $T_1$ and $T_2$ describing
 linear photocurrent in the following form ($\omega\tau \gg 1$)
\begin{subequations}
\label{t:qnt}
\begin{equation}
 T_1 = -e^3v^4 \frac{64\pi}{c\hbar\omega^4} \sum_{\bm k} [f(\varepsilon_k)
 - f(\varepsilon_{ p})]\frac{\varepsilon_p }{(\varepsilon_k + \varepsilon_p)^2}, \label{t1:qnt}
 \end{equation}
 \begin{equation}
T_2 = -e^3v^4 \frac{16\pi}{c\hbar\omega^4} \sum_{\bm k} [f(\varepsilon_k)
- f(\varepsilon_p)] \frac{\varepsilon_p^2 +\varepsilon_k^2 + (\hbar\omega)^2}{\varepsilon_k(\varepsilon_k + \varepsilon_p)^2}.\label{t2:qnt}
 \end{equation}
\end{subequations}
Here $\varepsilon_{p} = \varepsilon_{k} + \hbar\omega$. It is
noteworthy that Eqs.~\eqref{t:qnt} are valid provided $\hbar
\omega <E_F$. We note that although the scattering rates are not
explicitly present in Eqs.~\eqref{t:qnt}, the scattering processes
are crucial for the photocurrent formation.

If the photon frequency becomes much smaller as compared with the electron energies,
$\hbar\omega \ll \varepsilon_k, \varepsilon_p$, but $\omega\tau_1,\omega\tau_2 \gg 1$
the photon drag effect can be described classically. One can check that, in agreement with Eqs.~\eqref{hfBC}, Eqs.~\eqref{t:qnt} yield
\begin{equation}
\label{high-frq}
T_1 = 2T_2 = \frac{16\pi e^3v^4}{c\omega^3} \sum_{\bm k} \frac{ f'}{\varepsilon_{k}},
\end{equation}
where $f'=df/d\varepsilon$. In this frequency range values of
$T_1$ and $T_2$ are identical to those presented
in~\cite{condmat2010}. Hence, linear photocurrents   $j_B, j_C
\propto 1/\omega^2$ in this frequency range, see Eq.~\eqref{hfBC}.
Moreover, it can be shown that the circular high frequency Hall
effect requires an allowance for the extra scattering and
$\tilde{T}_1 \propto 1/\omega^4$ making $j_A \propto 1/\omega^3$
in agreement with Eq.~\eqref{hfA}. Therefore, the frequency
dependence of the circular photocurrent, $j_A$, is non-monotonous
with the maximum at $\omega\tau \sim 1$. This is exactly the
behavior observed experimentally, see Fig.~\ref{fig2bis}, where
the coefficient $A$ is plotted. Its absolute value first increases
with the frequency and afterwards rapidly decreases.
Overall agreement of the experimental data in sample 2 (shown by
the points) and theoretical calculation (solid line) shown in
Fig.~\ref{fig2bis} is good. The theory, however, does not describe
the abrupt frequency dependence and change of the photocurrent's
sign observed in sample 1 (see gray circles in
Fig.~\ref{fig2bis}). In order to understand this behaviour we
analyze the possible contributions of photogalvanic effects.

\subsubsection{Microscopic mechanisms of photogalvanic effects}

Real graphene samples are deposited on substrates. As we already
noted above, it results in a lack of an inversion center and,
correspondingly, allows for the photogalvanic effects.
Phenomenological analysis demonstrated that the polarization and
incidence angle dependences of the photogalvanic current are
almost the same as for the \emph{ac} Hall effect.  It follows from
the general arguments and phenomenological considerations
summarized in Eqs.~\eqref{j:pge}, that the photocurrent can be
generated only with allowance for $z$-component of the incident
electric field. However, for strictly two-dimensional model where
only $\pi$-orbitals of carbon atoms are taken into account, no
response at $E_z$ is possible. Therefore, microscopic mechanisms
of the photogalvanic effects in graphene involve other bands in
electron energy spectrum formed from the $\sigma$-orbitals of
carbon atoms.

\begin{figure}[hptb]
\includegraphics[width=0.6\linewidth]{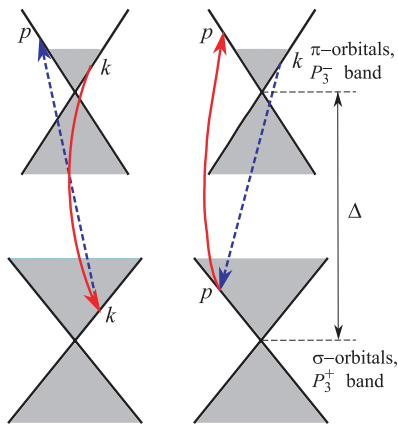}
\caption{Schematic illustration of indirect intraband transitions
with intermediate states in $P_3^+$ band which interfere with
Drude transitions (shown in Fig.~\ref{fig:inter}) and give rise to
the photogalvanic effect. Solid red arrows are electron-photon
interaction, dashed blue arrows are the electron scattering.}
\label{fig:inter1}
\end{figure}

There are 6 irreducible representations $P_1^+$, $P_1^-$, $P_2^+$,
$P_2^-$, $P_3^+$, and $P_3^-$  at $K$ (or $K'$) point of the
graphene's Brillouin zone.  The conduction and valence band states
transform according to the $P_3^-$ representation: there are two
basis functions $p_z^{(1)}$, $p_z^{(2)}$ being odd at the
reflection in the graphene plane $z=0$. Symmetry
analysis~\cite{Bassani_old,Bassani} shows that the transitions in
$z$ polarization are possible between these states (transforming
according to $P_3^-$) and the states transforming according to
$P_3^+$. The latter representation is described by two functions
$s^{(1)}$ and $s^{(2)}$ which do not change their signs at the
mirror reflection $z\to -z$. Under the symmetry operations which
do not involve $z\to -z$ these wave functions transform like
$p_z^{(1)}$, $p_z^{(2)}$. Representation $P_3^+$ corresponds to
$\sigma$ orbitals of carbon atoms which form remote valence and
conduction bands of graphene. Microscopic calculations performed
within the basis of $2s$ and $2p$ atomic
orbitals~\cite{Bassani_old,Bassani,Zunger} show that the distance
from the $P_3^-$ states forming conduction and valence bands and
closest deep valence bands $P_3^+$, $\Delta$, is about $10$~eV. It
is remarkable, that the electron dispersion in these bands has the
form, similar to that of conduction and valence bands: i.e. energy
spectrum near $K$ (or $K'$) point is linear, however, with
different velocity, as it is schematically illustrated in
Fig.~\ref{fig:inter1}.

Microscopically, circular photogalvanic effect arises due to the
quantum interference of the Drude transitions represented in
Fig.~\ref{fig:inter} (for $\bm q=0$) and the indirect intraband
transitions with intermediate states in $P_3^+$ bands depicted in
Fig.~\ref{fig:inter1}, similarly to the orbital mechanisms of the
photogalvanic effects in conventional semiconductor
nanostructures~\cite{Tarasenko2007,PhysRevB.79.121302,tarasenko11}.
Indeed, matrix elements of Drude transitions are proportional to
the in-plane components of electric field $\bm E_\parallel$ and
electron in-plane wave vectors in the initial $\bm k$, and final
$\bm p$ states. The matrix elements of the indirect transitions
via $P_3^+$ band are proportional to $E_z$ and do not contain
linear in $\bm k$, $\bm p$ contributions. As a result, the
interference contribution to the transition rate is proportional
to both $\bm E_\parallel$ and $E_z$ and to the in-plane wave
vector components giving rise to \emph{dc} current. The presence
of the substrate allows electron scattering between the states
transforming according to $P_3^+$ and $P_3^-$ representations: for
instance, the impurities located near the substrate surface or the
phonons, propagating in the substrate, or the impurities adsorbed
from the air to the graphene create an effective potential which
is not symmetric with respect to $z\to -z$ mirror reflection.
Hence, the interference contribution to the transition rate is
non-vanishing.

Let us denote $\sigma$ orbital states transforming according to
$P_3^+$ orbitals as $+'$ and $-'$ (we recall that the superscripts $+$
and $-$ denote the conduction and valence band states in
Eq.~\eqref{m:second}, respectively). We assume that the relevant interband
optical matrix element has a form~\cite{note1}
\begin{equation}
\label{Rinter}
R^{+'+}_{\bm k \bm k} = - R^{++'}_{\bm k \bm k} =- \frac{e}{m_0c} A_z \mathrm i p_{0},
\end{equation}
where $\mathrm i p_0$ is the momentum matrix element between $\sigma$
and $\pi$ orbitals, $p_0$ is assumed to be real (and the momentum
matrix element is imaginary).

We also need to define  the form of the  interband scattering matrix
elements. We have already noted that the phonons in the substrate
or the impurities positioned either above or below the graphene sheet
can provide the scattering between the bands transforming by $P_3^+$
and $P_3^-$ representations. In addition, the impurities or phonons should also
provide the scattering  within the $\pi$-orbital band. Such a scattering should be short-range
in order to allow the electron transition between $\sigma$ and $\pi$
orbitals. We assume that the interband scattering also takes place
between the similar combinations of the Bloch functions. We take the
scattering matrix elements in the following form for the  interband
scattering for the relevant processes~\cite{note1}:
\begin{equation}
\label{Vinter}
V^{+'+}_{\bm p \bm k} = V^{++'}_{\bm p \bm k} = \frac{V_1}{2} \left[1+
  e^{\mathrm i(\varphi_{\bm k} - \varphi_{\bm p})}\right],
\end{equation}
with $V_1$ being the real constant.

The second-order matrix element for the scattering-assisted optical
transition via $\sigma$ orbital can be written as
\begin{equation}
\label{Ms}
M^{\sigma}_{\bm p \bm k} = \frac{V^{++'}_{\bm p \bm k} R^{+'+}_{\bm k \bm k}}{\varepsilon_{ k,+}
- \varepsilon_{ k,+'}+ \hbar\omega}+ \frac{R^{++'}_{\bm p \bm p} V^{+'+}_{\bm p \bm k} }{\varepsilon_{ k,+} - \varepsilon_{ p,+'}}.
\end{equation}
Here $\varepsilon_{k,\nu}$ with $\nu=+$ or $+'$ describes electron
dispersion in a given band.
Corresponding processes are depicted in Fig.~\ref{fig:inter1}. To
simplify the calculations we assume that the dispersions of electron
in $\sigma$ and $\pi$ bands are the same. The allowance for difference
of effective velocities will result in the modification of the results
by the factor $\sim 2$. Equation~\eqref{Ms} under assumption  that
$\Delta \gg \hbar\omega, E_F$ transforms to
\begin{equation}
\label{Ms1}
M^{\sigma}_{\bm p \bm k}  \approx \mathrm i \frac{e A_z p_{0} V_1}{2m_0c}
\left[1+ e^{\mathrm i(\varphi_{\bm k} - \varphi_{\bm p})}\right] \frac{2\hbar\omega}{\Delta^2}.
\end{equation}

It is the quantum interference of the transitions via $\sigma$
orbitals described by Eq.~\eqref{Ms1} and Drude transitions
described by Eq.~\eqref{m:second} (where one has to put $\bm q
=0$)~\cite{Tarasenko2007,PhysRevB.79.121302} that gives rise to
the photocurrent. The photocurrent density under the steady-state
illumination can be written as [cf. Equation~\eqref{j:quant} and
Ref.~\cite{Tarasenko2007}]
\begin{multline}
\label{current} \bm j = e\frac{8\pi}{\hbar} \sum_{\bm k, \bm p}
2\Re{\left\{M^{\bm q=0}_{\bm p \bm k} M^{\sigma,*}_{\bm p \bm k}
\right\}} [\bm v_{\bm p} \tau_1(\varepsilon_{ p}) - \bm v_{\bm k}
\tau_1(\varepsilon_{ k})] \times \\
 [ f(\varepsilon_k) - f(\varepsilon_p)] \delta(\varepsilon_{ p} - \varepsilon_{ k} - \hbar\omega).
\end{multline}

Making necessary transformations we arrive at the following expression
for the constant $\chi_c$ describing circular photogalvanic effect:
\begin{multline}
\label{jy}
\chi_c = -ev \frac{4\pi w}{\hbar}  \sum_{\bm k \bm p}
\frac{\tau_1(\varepsilon_{ p}) \varepsilon_{ k}
  +\tau_1(\varepsilon_{ k})\varepsilon_{ p}}{\varepsilon_{ k}
  + \varepsilon_{ p}} \times \\ [ f(\varepsilon_k)
  - f(\varepsilon_p)] \delta(\varepsilon_{ p} - \varepsilon_{ k} - \hbar\omega),
\end{multline}
where
\[
w=\frac{2 {\pi} e^2 v p_0}{m_0 c\omega^2} \frac{\langle V_0V_1\rangle}{\Delta^2},
\]
and $\langle \ldots \rangle$ denote the averaging over disorder realizations.
Equation~\eqref{jy} is valid provided $\omega\tau \gg 1$ and
$\hbar\omega< E_F$. The treatment of the general case is given in
Appendix to the paper.

The direction of the current is determined by the
sign of the product $\langle V_0V_1\rangle$ and the radiation
helicity. The averaged
product $\langle V_0V_1\rangle$ has different signs for the same
impurities, but
positioned on top or bottom of graphene sheet. It is clearly seen that
the photogalvanic current vanishes in symmetric graphene-based
structures where $\langle V_0V_1\rangle=0$.

In the case of the degenerate electron gas with the Fermi energy $E_F$ and in the
limit of $\hbar \omega \ll E_F$ Eq.~\eqref{jy} can be recast as
\begin{equation}
\label{jy1}
\chi_c = -8\frac{\alpha e d_0 }{\Delta}  \frac{\langle V_0V_1\rangle}{\langle V_0^2\rangle} \frac{E_F}{\hbar \omega},
\end{equation}
we introduced effective dipole of interband transition
\[
e d_0 = \frac{e p_0 \hbar}{m_0 \Delta}.
\]
In Eq.~\eqref{jy1} $\alpha$ is the fine structure constant. It follows
from Eq.~\eqref{jy1} that the circular photocurrent caused by the
photogalvanic effect behaves as $1/\omega$ at $\omega\tau \gg 1$,
$\hbar\omega\ll E_F$, i.e. it is parametrically larger than the
circular ac Hall effect which behaves as $1/\omega^3$, see
Eq.~\eqref{hfA}. This important properly is related with the
time reversal symmetry: the coefficient $\chi_{c}$ describing
photogalvanic effect is even at time reversal while $\tilde T_1$
describing caHE is odd. Therefore, circular photocurrent formation
due to photogalvanic effect is possible at the moment of
photogeneration of carriers, making extra relaxation processes
unnecessary.

As discussed above experimental proof for the CPGE comes from
spectral sign inversion of the total photocurrent  observed in
sample~1  [see Figs.~\ref{fig2bis}, \ref{fig3} and
\ref{fig4}(a)]. Let us estimate the circular photocurrent
and compare it to experiment assuming that the photocurrent in sample~1 is dominated by the CPGE.
Taking $d_0 =1$~\AA, $\Delta = 10$~eV we obtain
\begin{equation}
\label{est}
\chi_c = A \sim   \frac{\langle V_0V_1\rangle}{\langle V_0^2\rangle}
\frac{ E_F}{\hbar \omega} \times 1.4 \times 10^{-11} \frac{\mbox{A cm}}{\mbox{W}} ,\quad \frac{\hbar}{\tau} \ll \hbar\omega \ll E_F.
\end{equation}
In the studied frequency range of CO$_2$ laser operation
$E_F/(\hbar\omega) \approx 3$. Considering the strongly asymmetric
scattering, where $\langle V_0V_1\rangle/{\langle V_0^2\rangle}
\approx 0.5$, our estimation yields  $A  \approx 2\times 10^{-11}$
(A cm)$/$W which is in a good agreement with experiment [see
Figs.~\ref{fig2bis}]. The values of the circular photocurrent
driven  by the PGE and by CaCHE are similar for $\hbar\omega \sim
100$~meV. It means, that for lower frequencies, the CaCHE
dominates, since it has stronger frequency dependence, while for
higher frequencies, the circular photogalvanic effect may take
over. While the sign of the circular ac Hall effect is determined
solely by the conductivity type in the sample and the radiation
helicity, the circular photogalvanic current sign depends on the
type of the sample asymmetry. In general, these two effects may
have opposite signs which may result in the sign inversion
observed in experiment, Fig.~\ref{fig2bis}.

The strongly asymmetric scattering might be exactly the case for
the short range impurities positioned on the substrate surface or
adsorbed from the air on the open surface of the sample and which
provide the same efficiency of both inter- and intra-band
scattering. Obviously, the degree of asymmetry and even its sign,
which reflects the coupling of the graphene layer with the
substate, depend on the growth conditions and may vary from sample
to sample. This explains the fact that the sign inversion is
detected only in some studied samples.

\section{Discussion and conclusions}

To summarize, we have carried out the detailed experimental
investigation of the photocurrents in graphene in the long
wavelength infrared range. The photocurrents were excited by
pulsed CO$_2$ laser at oblique incidence in large area epitaxial
graphene samples. The magnitudes and directions of the
photocurrents depend on the radiation polarization state and, in
particular, the major contribution to the photocurrent changes its
sign upon the reversal of the radiation helicity.

Phenomenological and microscopic theory developed in this work
show that there are two classes of effects being responsible for
the \emph{dc} current generation driven by polarization of the
radiation. Firstly, the photocurrent may arise due to the joint
action of the electric and magnetic fields of the electromagnetic
wave (or transfer of the radiation wave vector to the electron
ensemble). Secondly, the current may be generated due to the
photogalvanic effects which become possible when the inversion
symmetry is broken by the presence of the substrate. In this case,
the magnetic field of the radiation or its wave vector are not
important, but the asymmetry of the structure is needed. Arguments
based on the symmetry to the time reversal show that even in the
case of small asymmetry of the sample, the circular photogalvanic
effect can become parametrically dominant at high frequency due to
weaker decrease with an increase of the frequency ($1/\omega$ as
compared with $1/\omega^3$ for CacHE). While both types of
photocurrents are indistinguishible on the phenomenological level,
investigation of their  frequency dependence allowed to
distinguish them and provided direct experimental proof for the
existence of CPGE in graphene. Microscopic theory of $ac$ Hall
effect and CPGE give a good qualitative as well as quantitative
agreement of the experiment.

Our experiments also demonstrated that photocurrent exhibits
resonance behaviour at frequency close to edge of the reststrahlen
band of the SiC substrate at about $\hbar\omega \approx 121$~meV.
The resonance is observed for all photocurrent contributions and
may indicate an importance of the graphene coupling to the
substrate and role of the phonons in the substrate. The origin of
the resonance remains unclear and determination is a task of
future work.

\acknowledgements

We thank E.~L.~Ivchenko, S.~A.~Tarasenko, V.~V.~Bel'kov, D.~Weiss
and J.~Eroms for fruitful discussions and support. Support from
DFG (SPP~1459 and GRK~1570), DAAD, Linkage Grant of IB of BMBF at
DLR, Applications Center ''Miniaturised Sensorics'', Swedish
Research Council, SSF, RFBR, Russian Ministry of Education and
Sciences and ``Dynasty'' Foundation ICFPM is acknowledged.

\appendix

\section{Photogalvanic effects in classical frequency range}


In the case of $\hbar\omega \ll  E_F$ photogalvanic effects allow
a simple and physically transparent
interpretation~\cite{tarasenko11}: in asymmetric structures $z$
component of the incident electric field gives rise to the
temporal oscillations of the electron momentum scattering time
$\tau_1(t)$. As a result, $dc$ current is formed
\[
 \bm j \propto \overline{\tau_1(t) \bm E_\parallel(t)},
\]
where overline denotes temporal averaging.

The method developed in Ref.~\cite{tarasenko11} can be generalized for
graphene. Indeed, the processes depicted in Fig.~\ref{fig:inter1} and
described by the matrix element~\eqref{Ms1} can be interpreted as the
$E_z$ induced correction to the electron
scattering. Equation~\eqref{Ms1} can be recast as:
\begin{equation}
\label{Ms11}
M^{\sigma}_{\bm p \bm k} = \frac{e E_z(t) p_{0} V_1}{2m_0} \left[1+ e^{\mathrm i(\varphi_{\bm k} - \varphi_{\bm p})}\right] \frac{2\hbar}{\Delta^2}.
\end{equation}
As a result, the correction to the electron momentum scattering rate is given by
\begin{multline}
 \label{delta:tau}
\delta \left(\frac{1}{\tau}\right) =\\
 \frac{2\pi}{\hbar} \sum_{\bm p} 2\Re{\left[M^{\sigma}_{\bm p \bm k}
     (V^{++}_{\bm p \bm k})^*\right]}\delta(\varepsilon_{\bm p} -
 \varepsilon_{\bm k}) [1-\cos{(\varphi_{ p} - \varphi_{ k})}] =\\
 \zeta e E_z(t),
\end{multline}
where
\begin{equation}
 \zeta = S\frac{\langle V_0V_1\rangle}{v^2} \frac{d_0}{\hbar} \frac{\varepsilon_{ k}}{\Delta},
\end{equation}
where $S$ is the sample area.

Following Ref.~\onlinecite{tarasenko11} we obtain the photocurrent
density in the following form:
\begin{multline}
\label{j:class}
 \bm j =
 -\frac{8\alpha e d_0 \varepsilon_F}{\hbar \Delta}   \tau
 \frac{\langle V_0V_1\rangle}{\langle V_0^2\rangle}   I \times\\
\left[\frac{\bm e_\parallel e_z^*+\bm e_\parallel^*e_z}{1+(\omega\tau)^2} + \mathrm i (\bm e_\parallel e_z -  \bm e^*_\parallel e_z) \frac{\omega\tau}{1+(\omega\tau)^2} \right].
\end{multline}
In agreement with symmetry considerations, Eq.~\eqref{j:pge}, both linear and circular
photocurrents are allowed. For $\omega\tau \gg 1$ Eq.~\eqref{j:class} agrees with Eq.~\eqref{jy1}.


\begin{thebibliography}{99}
\bibitem{note1} {It is assumed that the
photon energy $\hbar\omega\leqslant E_F$ therefore both initial and final
states in the Drude transition take place in the same band, say, in
conduction band for $n$-type system. Hence, we need only this
matrix element}
\end{thebibliography}
\end{document}